\documentclass[aps,pra,twocolumn,floatfix]{revtex4-2}
\usepackage{amsmath,amssymb,graphicx,bbold}
\usepackage{graphicx}  
\usepackage{dcolumn}   
\usepackage{bm}        
\usepackage{color}
\usepackage{xcolor}

\usepackage{xcolor}
\usepackage{physics}
\usepackage{ulem}
\usepackage{subfigure}

\begin{document}

\title{Generalized Orbital Angular Momentum Symmetry in Parametric Amplification}
%




\author{R. B. Rodrigues, G. B. Alves, R. F. Barros, C. E. R. Souza, A. Z. Khoury}
\affiliation{Instituto de F\'{i}sica, Universidade Federal Fluminense, 24210-346 Niter\'{o}i, RJ, Brazil}
\date{\today}

\begin{abstract}
We investigate interesting symmetry properties verified by the down-converted beams produced in optical parametric amplification with structured light. We show that the Poincar\'e sphere symmetry, previously demonstrated for first-order spatial modes, translates to a multiple Poincar\'e sphere structure for higher orders. Each one of these multiple spheres is associated with a two-dimensional subspace defined by a different value of the orbital angular momentum. Therefore, the symmetry verified by first order modes is reproduced independently in each subspace. This effect can be useful for parallel control of independently correlated beams.
\end{abstract}

\maketitle
\section{Introduction}

Optical parametric amplification is a powerful tool for generating quantum correlations between independent light beams \cite{Reynaud1987,fabre1987,Mertz1991,pereira1992,villar2006,Cassemiro2007}. 
It has been used as an important resource for many quantum applications such as quantum teleportation \cite{Lam2003} and quantum metrology \cite{Treps2003}. The longitudinal-mode structure of quantum correlated beams generated by an optical parametric 
oscillator gives rise to a frequency comb of quadrature entangled beams that are good candidates to scalable quantum 
computers \cite{Pfister2008,Pfister2011,Pfister2014}.
These useful correlations stem from different constraints imposed by the parametric process, which includes energy and momentum conservation among the photons participating in the nonlinear interaction. Transverse momentum conservation is in the heart of well established spatial correlations between the photons 
emitted by spontaneous parametric down-conversion \cite{Monken1998}. These spatial correlations can be combined with polarization 
entanglement \cite{kwiat1999}, giving rise to hyperentangled two-photon quantum states
\cite{Kwiat1997,Santos2001,Pardal2003,kwiat2010}. 

Interesting conditions are also verified when structured light beams are coupled in the parametric process. Orbital angular momentum (OAM) conservation has been investigated in cavity-free spontaneous \cite{Mair2001} and stimulated \cite{caetano2002} parametric down-conversion. 
The nonlinear coupling between different transverse modes is subject to conditions imposed by the spatial overlap between them, giving rise to selection rules that limit the modes allowed in the interaction 
\cite{Buono_2014,Pereira2017,Buono:2018,sonja2021}. 
When the process is intensified inside an optical resonator, cavity conditions also dictate which modes can survive the loss-gain balance, which can affect both the transverse \cite{Lugiato1994,Marte1998,Schwob1998} and longitudinal \cite{Barros2021} mode structure. 
These effects determine whether OAM can be exchanged between the 
interacting modes \cite{martinelli2004,coutinho2008,alves2018,Aadhi2017}.
The selection rules that apply to parametric amplification also lead to symmetry properties that have already been investigated for first-order modes. In particular, OAM conservation and intensity overlap between the down-converted beams were shown to impose a reflection symmetry in the Poincar\'e sphere representaion of the signal and idler beams \cite{courtial1999,coutinho2007,belas2018,Oliveira2019}. 
OAM correlations inside an OPO give rise to 
entanglement in the continuous variable regime \cite{Andersen2009}, that can be combined with polarization to produce 
continuous variable hyperentanglement \cite{coutinho2009,gao2014,gao2018,lorenzo2019,gao2020}.

In this work, we investigate how this Poincar\'e sphere symmetry extends to higher orders. In principle, this subject suggests a difficult task, since higher-order modes do not have a simple geometric representation. However, the selection rules that arise from the spatial overlap between the interacting modes impose restrictions that limit the symmetry properties to two-dimensional subspaces of the higher order mode structure. These subspaces are spanned by pairs of modes with opposite OAM values. The Poincar\'e symmetry is independently verified inside each subspace, what can be useful for parallel control of independent down-conversion channels. Here, we will focus on the classical behaviour of the mode dynamics, which will serve as a starting point for a future investigation in the quantum domain. As we will see, this classical instance of the problem already encompasses a rich dynamics.

\section{Structured light injection in parametric amplification}
\label{section:dyneq}

\begin{figure}
	\includegraphics[scale=0.14]{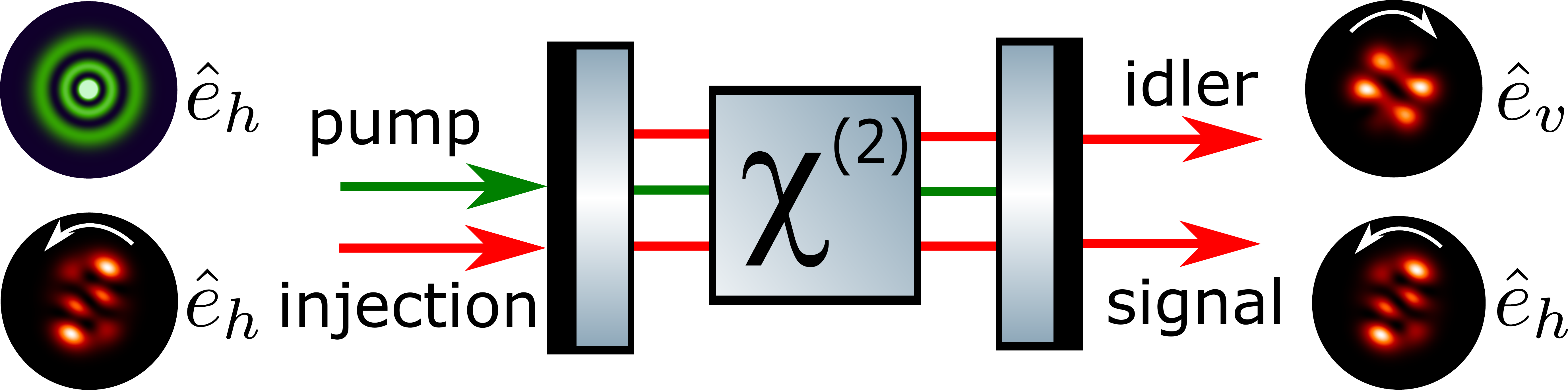}
	\caption{Sketch of the optical parametric oscillator with type-II phase-matching and structured light injection. $\hat{e}_h$ and $\hat{e}_v$ are the horizontal and vertical polarization unit vectors, respectively. The pump beam can be radially structured to 
	optimize the spatial overlap with signal and idler.}
	\label{fig:OPO_setup}
\end{figure}

Let us consider the optical parametric amplification process involving two input beams, \textit{pump} and \textit{signal}, which interact through a nonlinear crystal and generate a third beam called \textit{idler}. The interacting beams carry the frequencies $\omega_p$ (pump), $\omega_s$ (signal) and $\omega_i$ (idler), satisfying $\omega_p = \omega_s + \omega_i\,$. The three-beam interaction is mediated by the second order nonlinear susceptibility of the crystal. We are interested in deriving general symmetry properties carried by the signal and idler beams as a result of the nonlinear coupling. This kind of symmetry has already been investigated, both theoretical \cite{coutinho2007} and experimentally \cite{belas2018}, for first order modes, where OAM conservation and intensity overlap were the main features behind the symmetry observed. Our objective is to extend these symmetry properties to higher order modes injected in the OPO. The physical situation is illustrated in Fig. \ref{fig:OPO_setup}. A pump beam is sent to the OPO cavity along with a seed beam that matches the signal frequency and polarization. Inside the resonator, the pump energy is transferred to signal and idler, which, under type-II phase-matching, is generated with its polarization orthogonal to the signal beam. 

The seed beam is assumed to be structured with an arbitrary superposition of Laguerre-Gaussian (LG) modes of the same order $N_s$, while the pump beam is assumed to be in a single LG mode without OAM. In the LG basis, the pump and seed electric fields can be written as
\begin{eqnarray}
\mathcal{E}_{p_{in}} (\mathbf{r}) &=& \alpha_{p_{in}} \psi^{0q_p}(\mathbf{r})\;,
\nonumber\\
\mathcal{E}_{s_{in}} (\mathbf{r}) &=& \sum_{l}\alpha_{s_{in}}^{lq_s}\psi^{lq_s}(\mathbf{r})\;, 
\label{psi_in_geral}
\end{eqnarray}
where $\psi^{l q}(\mathbf{r})$ is a LG mode function with topological charge (OAM) $l$ and radial index $q\,$, 
$\alpha_{s_{in}}^{l q}$ is the corresponding complex amplitude of the signal mode and $\alpha_{p_{in}}$ is the 
pump complex amplitude. 
The mathematical expression of the LG modes in terms of the polar coordinates $(r,\phi)$ in the focal plane 
($z=0$) is given by \cite{siegman1986lasers}
\begin{equation}
\psi^{l q}(\mathbf{r}) = 
\sqrt{\frac{\left(2/\pi\right)\,p!}{\left(p+\abs{l}\right)!}}
\frac{\left(\sqrt{2}\,r\right)^{\abs{l}}}{w^{\abs{l}+1}} 
L_p^{\abs{l}}\left(\frac{2r^2}{w^2}\right) e^{-\frac{r^2}{w^2}}\, e^{il\phi},
\label{LGmodes}
\end{equation}
where $w$ is the beam radius and $L_p^{\abs{l}}\left(\frac{2r^2}{w^2}\right)$ is the 
generalized Laguerre polynomial.

The summation over the seed modes is constrained by $2q_s+\abs{l}=N_s\,$. The choice of a fixed order for the seed beam is of experimental relevance, since in this case all components evolve with the same Gouy phase and can be simultaneously mode matched to the OPO cavity.

\subsection{Dynamical Equations}
\label{subsec:dyneq}

These input modes feed the dynamics that governs the build up of the intracavity fields. They constitute the source terms of the dynamical equations for the intracavity amplitudes. 
Let the intracavity electric fields in the LG basis be written as
\begin{eqnarray}
\mathcal{E}_{p} (\mathbf{r}) &=& \alpha_{p} \psi^{0q_p}(\mathbf{r})\;,
\nonumber\\
\mathcal{E}_j(\mathbf{r}) &=& \sum_{l}\alpha_j^{lq_j}\psi^{lq_j}(\mathbf{r}), \label{psi_geral}   
\end{eqnarray} 
where the index $j=s,i$ refers to signal and idler, respectively. 
The stimulated idler beam will populate the Laguerre-Gaussian modes with optimal overlap 
\begin{eqnarray}
\Lambda^{l}_{q_sq_i} = \int \left[\psi^{0q_p}(\mathbf{r})\right]^\ast \psi^{lq_s}(\mathbf{r})\,\psi^{-lq_i}(\mathbf{r})\,d^2\mathbf{r}\;,
\end{eqnarray}
with the pump 
and seed modes. This imposes OAM conservation $l_s + l_i = l_p$ and restricts the radial indices as well. Since the pump 
beam is assumed to carry zero OAM, the coupled signal and idler modes must have opposite topological charges. However, the 
radial mode selection for the idler beam is not so simple. It is determined by the maximum overlap with the pump and seed modes.
This point will be clarified in our numerical examples.

Assuming the perfect resonance of the three fields, the dynamical equations for the 
intracavity mode amplitudes are
\begin{eqnarray}
\dot{\alpha}_p &=& -\gamma_p \alpha_p + i \chi \sum_{l} \Lambda_{q_sq_i}^{l}\, \alpha^{lq_s}_s\alpha^{-lq_i}_i
+\eta_p\alpha_{p_{in}}\;,  
\nonumber\\
\dot{\alpha}^{lq_s}_s &=& - \gamma \alpha^{lq_s}_s + i\chi\Lambda_{q_sq_i}^{l\,\ast}\, \alpha_p \left(\alpha^{-lq_i}_i\right)^\ast + 
\eta\alpha^{lq_s}_{s_{in}}  \;,
\label{dyneqs-0}\\
\dot{\alpha}^{lq_i}_i &=& - \gamma \alpha^{lq_i}_i + i\chi\Lambda_{q_sq_i}^{l\,\ast}\, \alpha_p \left(\alpha^{-lq_s}_s\right)^\ast\;,
\nonumber
\end{eqnarray}
where $\chi$ is the nonlinear coupling constant, $\gamma_p$ is the pump decay rate, $\gamma$ is the common decay rate of signal and idler, $\eta_p$ and $\eta$ are the pump and signal input transmissions, respectively.
We recall that the mode indices $l$ and $q_s$ run over the allowed values compatible with the seed order $N_s=2q_s+\abs{l}\,$, 
while the stimulated idler beam will carry the Laguerre-Gaussian modes with optimal overlap with the pump and seed modes. 

Our analysis is significantly simplified when we define the normalized variables
\begin{eqnarray}
\beta_p &=& \chi\alpha_p/\gamma \,, \quad \beta_{p_{in}} = \chi\eta_p\alpha_{p_{in}}/\gamma^2\,,
\nonumber\\
\beta^{lq_s}_s &=& \chi\alpha^{lq_s}_s/\gamma \,,
\quad
\beta^{lq_s}_{s_{in}} = \chi\eta\alpha^{lq_s}_{s_{in}}/\gamma^2\,,
\label{normvar}\\
\beta^{lq_i}_i &=& \chi\alpha^{lq_i}_i/\gamma \,.
\nonumber
\end{eqnarray}
With the normalized variables, the dynamical equations become 
\begin{eqnarray}
\dot{\beta}_p &=& -\gamma_r \beta_p + i \sum_{l} \Lambda_{q_sq_i}^{l}\, \beta^{lq_s}_s\beta^{-lq_i}_i
+\beta_{p_{in}}\;,  
\nonumber\\
\dot{\beta}^{lq_s}_s &=& - \beta^{lq_s}_s + i \Lambda_{q_sq_i}^{l\,\ast}\, \beta_p \left(\beta^{-lq_i}_i\right)^\ast + 
\beta^{lq_s}_{s_{in}}  \;,
\label{dyneqs}\\
\dot{\beta}^{lq_i}_i &=& - \beta^{lq_i}_i + i \Lambda_{q_sq_i}^{l\,\ast}\, \beta_p \left(\beta^{-lq_s}_s\right)^\ast\;,
\nonumber
\end{eqnarray}
where derivatives in the left-hand-side are taken with respect to the dimensionless time $\tau = \gamma t$ and 
we introduced the decay ratio $\gamma_r = \gamma_p/\gamma\,$.

\subsection{Steady State Solution}
\label{subsec:steady state}

The output field distribution is given by the steady state solution of the dynamical equations \eqref{dyneqs}, which can be obtained by 
setting the time derivatives equal to zero in the left-hand-side 
($\dot{\beta}_p=\dot{\beta}^{lq_s}_s=\dot{\beta}^{lq_i}_i=0$) 
and solving the resulting algebraic equations. From the last two equations we get
\begin{eqnarray}
\beta^{lq_s}_s &=& 
\frac{\beta^{lq_s}_{s_{in}}}{1-\abs{\Lambda^{l}_{q_sq_i}\,\beta_p}^2}  \;,
\nonumber\\
\beta^{lq_i}_i &=&   \frac{i\,\beta_p\,\Lambda^{l\,\ast}_{q_sq_i}\,\left(\beta^{-lq_s}_{s_{in}}\right)^\ast}
{1-\abs{\Lambda^{l}_{q_sq_i}\,\beta_p}^2}\;.
\label{ss-signal-idler}
\end{eqnarray}
These equations can be plugged into the steady state condition for the intracavity pump amplitude, resulting in
\begin{equation}
\left[\gamma_r + \sum_{l} \frac{\abs{\Lambda^{l}_{q_sq_i}\beta_{s_{in}}^{lq_s}}^2}
{\left(1-\abs{\Lambda^{l}_{q_sq_i}\,\beta_p}^2\right)^2}\right]\,\beta_p
= \beta_{p_{in}}\,,  
\label{ss-pump}
\end{equation}
where we used $\Lambda^{-l}_{q_sq_i} = \Lambda^{l}_{q_sq_i}\,$. 
Without loss of generality, we may set the input pump phase equal to zero 
($\alpha_{p_{in}}\in \mathbb{R}$). Therefore, the intracavity pump amplitude is also a real number that can be found by 
solving the quintic equation \eqref{ss-pump}. For arbitrary mode orders, this is usually a difficult task that is beyond 
the scope of this work. Nevertheless, Eqs. \eqref{ss-signal-idler} allow us to establish an interesting property of the 
down-converted beams generated by the nonlinear process, without the need of the intracavity pump solution. As we discuss 
next, the relationship between the amplitudes of the seed beam and the intracavity down-converted fields sets an interesting symmetry 
relation between signal and idler in a generalized Poincar\'e sphere representation of higher order modes.

\section{Generalized Poincar\'e Symmetry}
\label{sec:poincare}

The Poincar\'e sphere representation of OAM beams has been first introduced for first-order modes \cite{courtial1999}. 
It describes linear combinations 
of Laguerre-Gaussian modes with radial number $q=0$ and topological charges $l=\pm 1\,$. In our 
case, we will use an independent Poincar\'e sphere for each two-dimensional mode space spanned by Laguerre-Gaussian 
beams with opposite OAM, $\psi^{\pm l q}\,$. For the signal beam, $l$ and $q_s$ run over the allowed values 
compatible with $2q_s + \abs{l} = N\,$. For the idler beam, the radial numbers are defined by those modes with maximal spatial 
overlap with the pump and seed modes. 
Note that a zero OAM component $l=0$ can only occur in even orders for $q_s=N/2\,$, while the LG modes with odd orders 
have $l\neq 0\,$. In this way, we can group the LG modes of a given order in pairs with opposite OAM $\{\psi^{\pm l q}\}$ and 
an isolated mode with zero OAM for even orders. For example, for seed beams with orders from $0$ to $4$ we have
\begin{eqnarray}
&N=0:&\quad \{\psi^{0 0}\}\;,
\nonumber\\
&N=1:&\quad \{\psi^{\pm 1 0}\}\;,
\nonumber\\
&N=2:&\quad \{\psi^{\pm 2 0}\}\oplus\{\psi^{0 1}\}\;,
\nonumber\\
&N=3:&\quad \{\psi^{\pm 3 0}\}\oplus\{\psi^{\pm 1 1}\}\;,
\nonumber\\
&N=4:&\quad \{\psi^{\pm 4 0}\}\oplus\{\psi^{\pm 2 1}\}\oplus\{\psi^{0 2}\}\;.
\label{eq:mode-structure}
\end{eqnarray}
Note that each $\{\psi^{\pm \abs{l} q_s}\}$ subspace realizes an independent SU(2) structure.

The idler modes will follow a similar structure. However, the corresponding radial numbers are selected by the optimal overlap 
with the pump and seed modes and, in general, do not fix a given order. 
As we will see, the Poincar\'e sphere symmetry previously demonstrated for first order modes in Refs. \cite{coutinho2007,belas2018} is independently verified within each one of the two-dimensional OAM subspaces for higher orders.

From Eqs. \eqref{ss-signal-idler} we can see that the intracavity signal and idler amplitudes are related by 
\begin{equation}
\beta^{lq_i}_i = i\,\beta_p\, \Lambda_{q_sq_i}^{l\,\ast} \left(\beta^{-lq_s}_{s}\right)^\ast\;.
\label{signal-idler-relation}
\end{equation}
For $l=0\,$, there is no SU(2) structure and this equation simply states the conjugate relation between signal and idler amplitudes 
for the zero OAM modes. In this case, no Poincar\'e symmetry can be realized. However, when $l\neq 0\,$, Eq. \eqref{signal-idler-relation} sets a connection between the SU(2) structures of signal and idler. 
Let the signal input be an arbitrary structure of order $N=2q_s+\abs{l}\,$, which can be written as
\begin{equation}
    \mathcal{E}_{s_{in}} = \sum_{l > 0} A_{in}^{l q_s} 
    \left[\cos\left(\theta_{l}/2\right) \psi^{l q_s} + e^{i\phi_l} \sin\left(\theta_l/2\right) \psi^{-l q_s}\right],
\end{equation}
where $\{A_{in}^{l q}\}$ are complex amplitudes and $\{(\theta_l,\phi_l)\}$ are the Poincar\'e sphere 
coordinates that represent the seed mode in each SU(2) structure $\mathcal{H}^{l}\equiv\{\psi^{\pm l q}\}\,$.
For a given order $N\,$, a Poincar\'e sphere is associated with each OAM value $l\,$.
With these definitions, the source terms that figure in the dynamical equations \eqref{dyneqs} become
\begin{eqnarray}
    \beta^{l q_s}_{s_{in}} &=& \frac{\chi\eta}{\gamma^2}\,A_{in}^{l q_s}\,\cos\left(\theta_l/2\right) \;,
    \nonumber\\
    \beta^{-l q_s}_{s_{in}} &=& \frac{\chi\eta}{\gamma^2}\,A_{in}^{l q_s}\,e^{i\phi_l} \sin\left(\theta_l/2\right) \;.
    \label{signal-input-poincare-coordinates}
\end{eqnarray}
From the steady state solution \eqref{ss-signal-idler} and the signal-idler conjugation relation 
\eqref{signal-idler-relation}, we easily get
\begin{eqnarray}
    \beta^{l q_s}_{s} &=& \frac{\chi}{\gamma}\xi_{s}^{l q_s}\,\cos\left(\theta_l/2\right), \;
    \beta^{-l q_s}_{s} = \frac{\chi}{\gamma}\xi_{s}^{l q_s}\,e^{i\phi_l} \sin\left(\theta_l/2\right),
    \nonumber\\
    \beta^{l q_i}_{i} &=& \frac{\chi}{\gamma}\xi_{i}^{l q_i}\,e^{-i\phi_l} \sin\left(\theta_l/2\right),\;
    \beta^{-l q_i}_{i} = \frac{\chi}{\gamma}\xi_{i}^{l q_i}\,\cos\left(\theta_l/2\right),
    \nonumber\\
    \label{signal-idler-poincare-coordinates}
\end{eqnarray}
where
\begin{eqnarray}
    \xi_{s}^{l q_s} &=& \frac{\eta\,A_{in}^{l q_s}/\gamma}
    {1-\abs{\Lambda^{l}_{q_sq_i} \beta_p}^2}\,,
    \nonumber\\
    \xi_{i}^{l q_i} &=& i\,\beta_p\,\Lambda^{l\,\ast}_{q_sq_i}\,\left(\xi_{s}^{l q_s}\right)^\ast\,.
    \label{csi-signal-idler}
\end{eqnarray}
Equations \eqref{signal-idler-poincare-coordinates} set the Poincar\'e sphere symmetry between signal and idler spatial
modes. Indeed, we can easily see that signal and idler coordinates on the sphere are related by
\begin{eqnarray}
    \theta_l^i &=& \pi - \theta_l^s\;,
    \nonumber\\
    \phi_l^i &=& \phi_l^s\;,
\end{eqnarray}
which means that within each SU(2) structure $\mathcal{H}^{l}\,$, signal and idler are represented by two points 
on the sphere that are the specular image of each other with respect to the equatorial plane. This is a generalization 
of the first-order mode symmetry previously demonstrated in Refs. \cite{coutinho2007,belas2018}. The intracavity 
signal and idler spatial modes are then given by
\begin{eqnarray}
    \mathcal{E}_{s} &=& \sum_{l>0} \xi_{s}^{l q_s} 
    \left[\cos\left(\theta_{l}/2\right) \psi^{l q_s} + e^{i\phi_l} \sin\left(\theta_l/2\right) \psi^{-l q_s}\right] ,
    \nonumber\\
    \mathcal{E}_{i} &=& \sum_{l>0} \xi_{i}^{l q_i} 
    \left[e^{-i\phi_l} \sin\left(\theta_{l}/2\right) \psi^{l q_i} + \cos\left(\theta_l/2\right) \psi^{-l q_i}\right] .
    \nonumber\\
    \label{signal-idler-intracavity}
\end{eqnarray}
We next discuss some examples which allow us to visualize the generalization of the Poincar\'e sphere symmetry between 
signal and idler spatial modes. As we will see, the odd modes already capture the essential features of the symmetry, 
since for even orders the zero OAM components do not possess the required SU(2) structure. 

\subsection{First order revisited}
\label{subsec:1st-order}

We now briefly revisit the first order case already discussed in Refs. \cite{coutinho2007,belas2018}. In this case, 
we can assume a Gaussian pump ($q_p = 0$) and a first order signal input,
\begin{eqnarray}
    \mathcal{E}_{p_{in}} &=& \alpha_{p_{in}} \,\psi^{0 0}\,,
        \label{input-1st}\\
    \mathcal{E}_{s_{in}} &=& A_{in}^{1 0} 
    \left[\cos\left(\theta/2\right) \psi^{1 0} + e^{i\phi} \sin\left(\theta/2\right) \psi^{-1 0}\right]\,.
    \nonumber
\end{eqnarray}
The steady state intracavity pump amplitude is given by the solution of the quintic equation
\begin{equation}
\left[\gamma_r + \frac{\abs{\Lambda^{1}_{00}\,A_{in}^{10}}^2}
{\left(1-\abs{\Lambda^{1}_{00}\,\beta_p}^2\right)^2}\right] \beta_p 
= \beta_{p_{in}}\,.  
\label{ss-pump-1st-order}
\end{equation}
Then, the intracavity signal and idler spatial structures are 
\begin{eqnarray}
    \mathcal{E}_{s} &=& \xi_{s}^{1 0} 
    \left[\cos\left(\theta/2\right) \psi^{1 0} + e^{i\phi} \sin\left(\theta/2\right) \psi^{-1 0}\right]\,,
    \nonumber\\
    \mathcal{E}_{i} &=& \xi_{i}^{1 0} 
    \left[e^{-i\phi} \sin\left(\theta/2\right) \psi^{1 0} + \cos\left(\theta/2\right) \psi^{-1 0}\right]\,,
    \nonumber\\
    \label{signal-idler-intracavity-1st-order}
\end{eqnarray}
with the mode amplitudes given by
\begin{eqnarray}
    \xi_{s}^{1 0} &=& \frac{\eta\,A_{in}^{1 0}/\gamma}{1-\abs{\Lambda^{1}_{0 0} \beta_p}^2}\,,
    \nonumber\\
    \xi_{i}^{1 0} &=& i\beta_p \Lambda^{1\,\ast}_{0 0}\, \left(\xi_{s}^{1 0}\right)^\ast\,.
    \label{signal-idler-csi-1st-order}
\end{eqnarray}

\begin{figure}[h!]
	\includegraphics[scale=0.12]{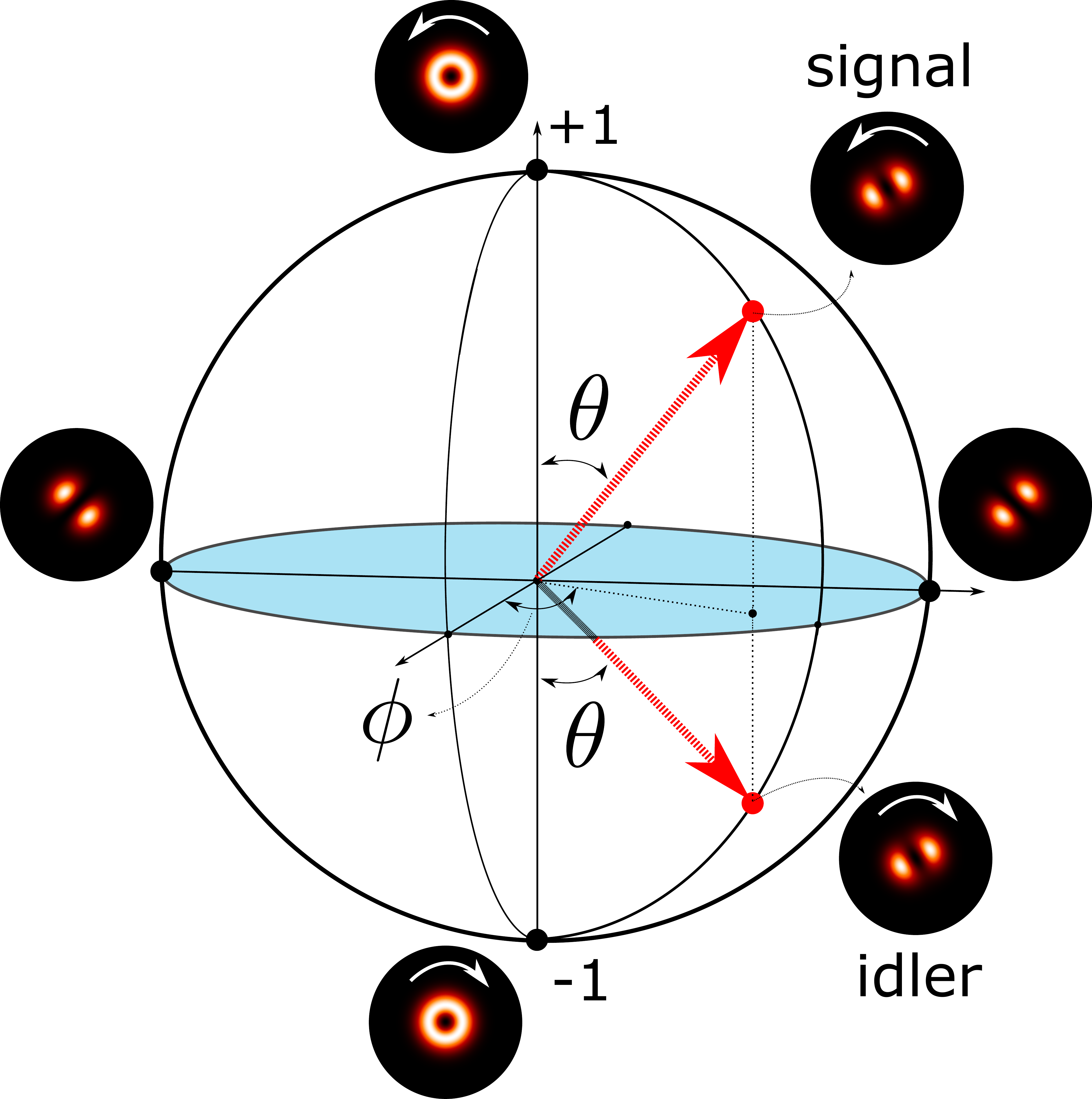}
	\caption{Poincaré sphere representation for first-order OAM symmetry.}
	\label{fig:simetria_esfera_2}
\end{figure}

The coordinates of the points representing the signal and idler structures in the Poincar\'e sphere are related by
\begin{eqnarray}
    \theta^i &=& \pi - \theta^s\;,
    \nonumber\\
    \phi^i &=& \phi^s\;.
\end{eqnarray}
As shown in Fig. \ref{fig:simetria_esfera_2}, 
the point representing the idler mode is the specular image of the point representing the signal 
with respect to the equatorial plane. This symmetry provides optimal intensity overlap and OAM 
conservation between signal and idler.

\subsection{Poincar\'e symmetry with a second order seed}
\label{subsec:2nd-order}

\begin{figure}[h!]
	\includegraphics[scale=0.27]{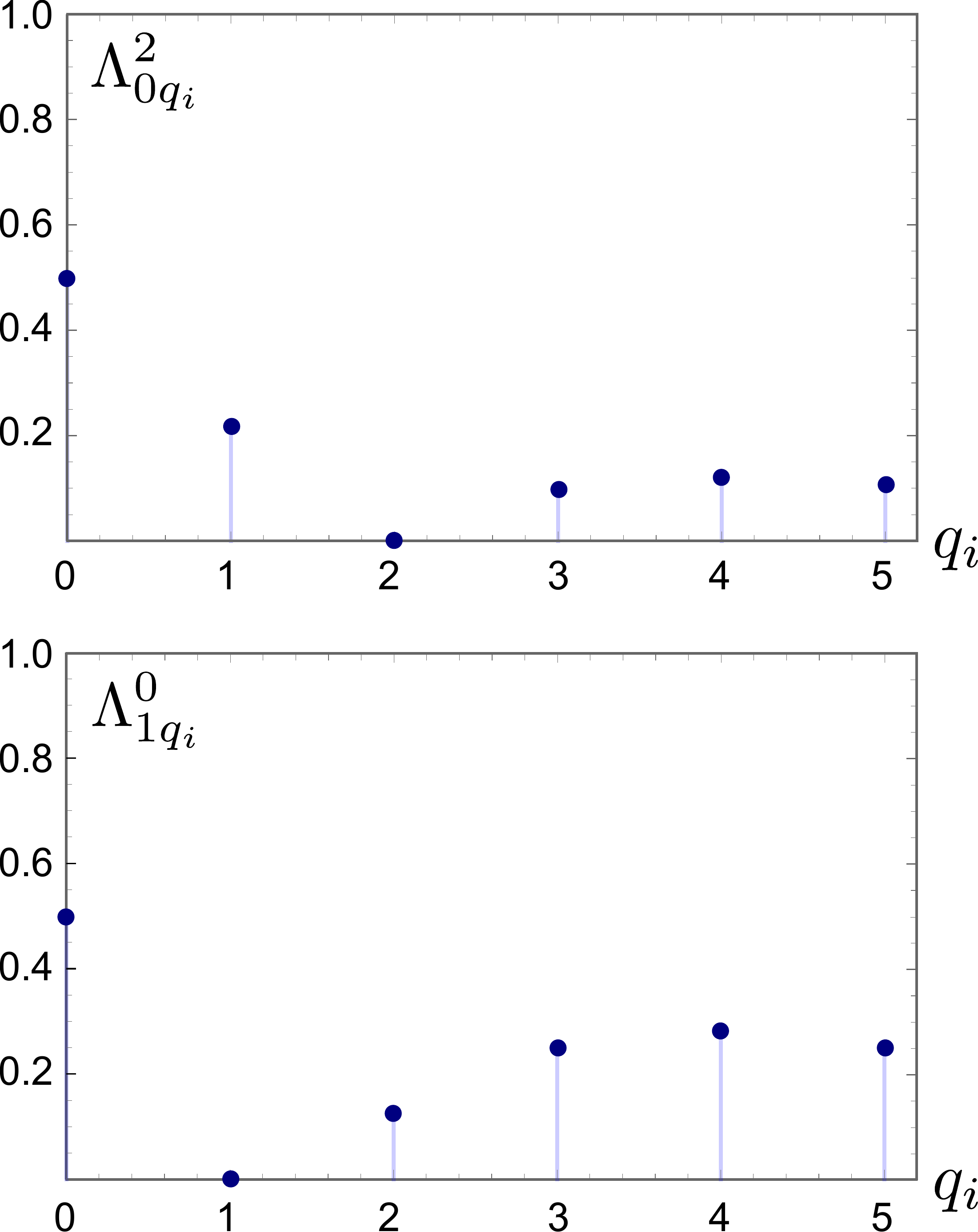}
	\caption{Pump-signal-idler overlap integrals for $l=\pm 2$ (top) and $l=0$ (bottom) as a function of the idler radial orders, when the pump parameters are fixed at $l_p = 0$ and $q_p = 1\,$.}
	\label{fig:overlap_2nd_order}
\end{figure}

The even-order subspaces include a zero-OAM mode with radial number $q=N/2\,$. Since it is an isolated single-mode 
subspace, there is no room for a Poincar\'e sphere representation or symmetry relation. The remaining OAM carrying 
modes can be grouped in pairs with opposite OAM, constituting a set of independent SU(2) structures where the 
aforementioned symmetry is verified. For example, consider the case of a second order seed beam 
$\{\psi^{\pm2 0},\psi^{0 1}\}$ and a single-mode pump with zero OAM and radial order $q_p=1\,$, 
\begin{eqnarray}
    \mathcal{E}_{p_{in}} \!&=& \alpha_{p_{in}} \,\psi^{0 1}\,,
    \label{input-2nd}\\
    \mathcal{E}_{s_{in}} \!&=&\! A_{in}^{0 1} \psi^{0 1} 
    \!+\! A_{in}^{2 0} 
    \left[\cos\left(\theta/2\right) \psi^{2 0} \!+\! e^{i\phi} \sin\left(\theta/2\right) \psi^{-2 0}\right] .
    \nonumber
\end{eqnarray}
\begin{figure}[h!]
	\includegraphics[scale=0.2]{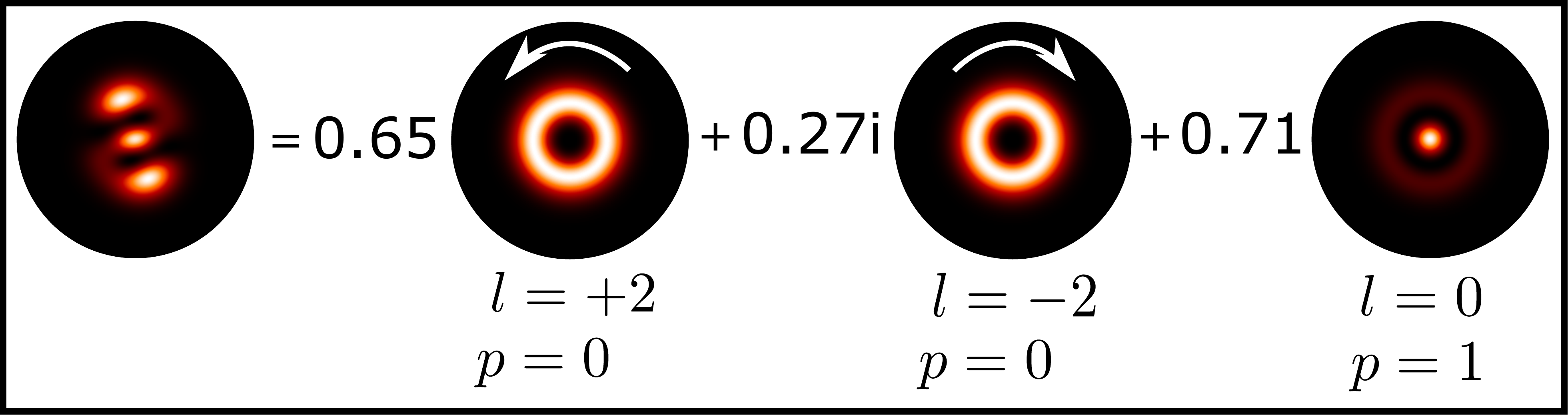}
	\caption{Example of a second-order seed beam. In this case we have set 
	$A_{in}^{0 1}=A_{in}^{2 0}=1/\sqrt{2}\,$, $\theta=45^o$ and $\phi=90^o$.}
	\label{fig:second_seed}
\end{figure}
An example of one such structure is shown in Fig. \ref{fig:second_seed}.
The idler modes which will profit from the pump and seed energy are those with maximum spatial overlap with the input modes. First, OAM conservation is required for non-vanishing overlap. 
Then, the radial order 
associated with each OAM is determined by the maximum numerical value of the overlap integrals 
$\Lambda^2_{0 q_i}$ and $\Lambda^0_{1 q_i}$. In Fig. \ref{fig:overlap_2nd_order} 
we show the numerical value of the overlap integrals as a function of the idler radial order. 

\begin{figure}[h!]
	\includegraphics[scale=0.12]{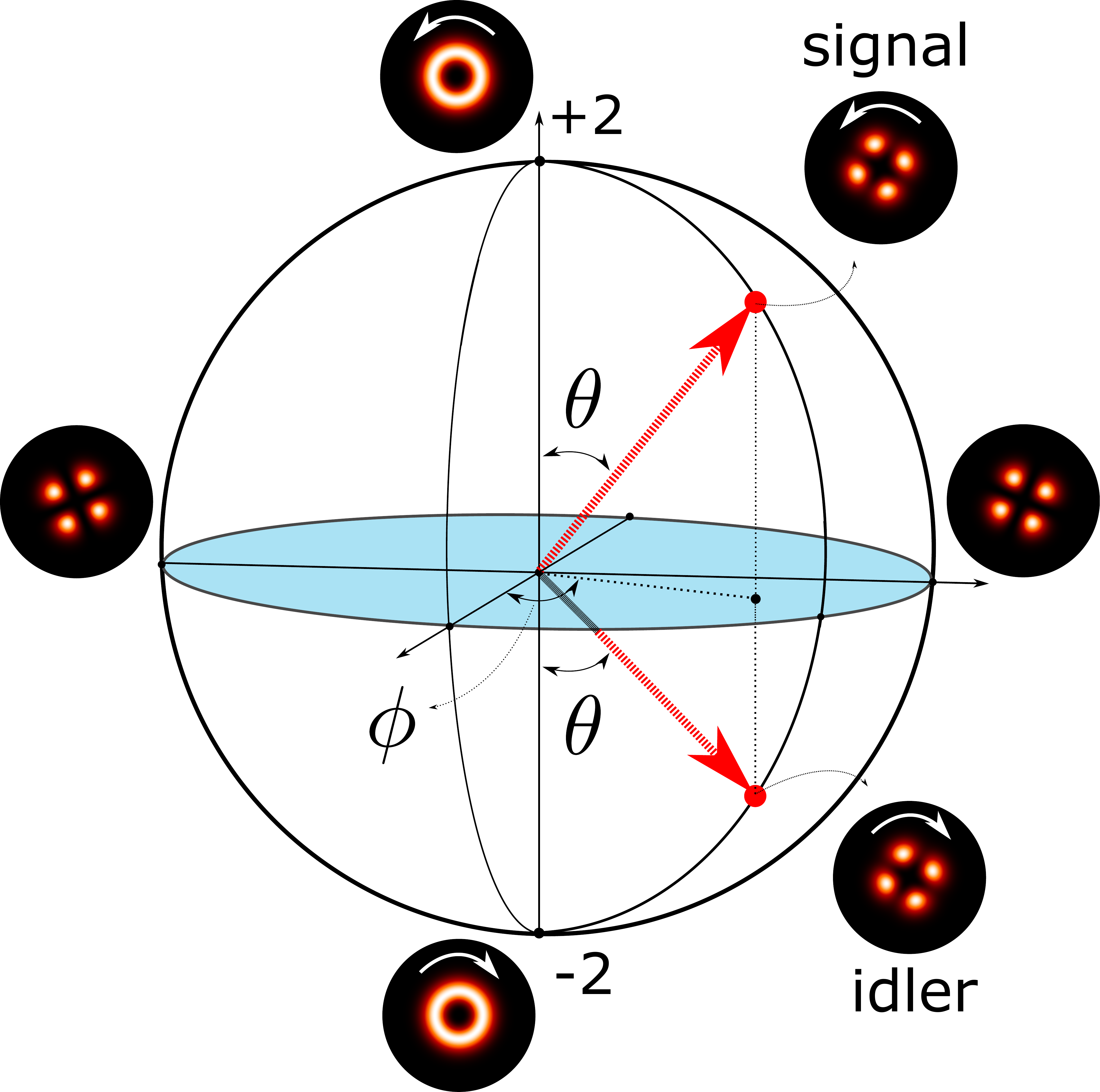}
	\caption{Signal and idler symmetry in the Poincar\'e sphere for 2nd order OAM ($l=\pm 2$).}
	\label{fig:spherep_2nd_order}
\end{figure}

As we can see, for both $l=0$ and $l=\pm 2\,$, the zero radial order ($q_i=0$) displays optimal 
coupling. Therefore, the 
transverse modes taking part in the intracavity interaction are $\{\psi^{0 1}\}$ for the pump, 
$\{\psi^{0 1},\psi^{\pm 2 0}\}$ for the signal and $\{\psi^{0 0},\psi^{\pm 2 0}\}$ for the idler. 
In this case, the pump steady-state is given by the solution of
\begin{equation}
\left[ \gamma_r +
\frac{\abs{\Lambda^{0}_{10}\,A_{in}^{01}}^2}
{\left(1 \!-\! \abs{\Lambda^{0}_{10}\,\beta_p}^2\right)^{\!2}}
+
\frac{\abs{\Lambda^{2}_{00}\,A_{in}^{20}}^2}
{\left(1 \!-\! \abs{\Lambda^{2}_{00}\,\beta_p}^2\right)^{\!2}}
\right] \beta_p = \beta_{p_{in}}\,,
\label{ss-pump-2nd-order}
\end{equation}
and the steady state amplitudes of signal and idler are
\begin{eqnarray}
    \mathcal{E}_{s} &=& \xi_{s}^{0 1} \psi^{0 1} +
    \xi_{s}^{2 0} 
    \left[\cos\left(\theta/2\right) \psi^{2 0} + e^{i\phi} \sin\left(\theta/2\right) \psi^{-2 0}\right]\,,
    \nonumber\\
    \mathcal{E}_{i} &=& \xi_{i}^{0 0} \psi^{0 0} +
    \xi_{i}^{2 0} 
    \left[e^{-i\phi} \sin\left(\theta/2\right) \psi^{2 0} + \cos\left(\theta/2\right) \psi^{-2 0}\right]\,,
    \nonumber\\
    \label{signal-idler-intracavity-2nd-order}
\end{eqnarray}
with the mode amplitudes given by
\begin{eqnarray}
    \xi_{s}^{0 1} &=& \frac{\eta\,A_{in}^{0 1}/\gamma}{1-\abs{\Lambda^{0}_{1 0} \beta_p}^2}\, , 
    \quad 
    \xi_{s}^{2 0} = \frac{\eta\,A_{in}^{2 0}/\gamma}{1-\abs{\Lambda^{2}_{0 0} \beta_p}^2}\, , 
    \nonumber\\
    \xi_{i}^{0 0} &=& i\,\beta_p\,\Lambda^{0\,\ast}_{1 0}\,\left(\xi_{s}^{0 1}\right)^\ast , 
    \quad\!\!
    \xi_{i}^{2 0} = i\,\beta_p\,\Lambda^{2\,\ast}_{0 0}\,\left(\xi_{s}^{2 0}\right)^\ast .
    \label{signal-idler-csi-2nd-order}
\end{eqnarray}
Note that no special symmetry can be realized in the zero OAM subspace, only the usual conjugation relation 
between signal and idler amplitudes. However, as shown in Fig. \ref{fig:spherep_2nd_order}, 
the $l=\pm 2$ subspace displays the same 
kind of Poincar\'e sphere symmetry as the first order case, with the signal and idler coordinates related by 
\begin{eqnarray}
    \theta^i &=& \pi - \theta^s\;,
    \nonumber\\
    \phi^i &=& \phi^s\;.
\end{eqnarray}

\subsection{Two-sphere symmetry for third-order beams}
\label{subsec:3rd-order}

\begin{figure}[h!]
	\includegraphics[scale=0.27]{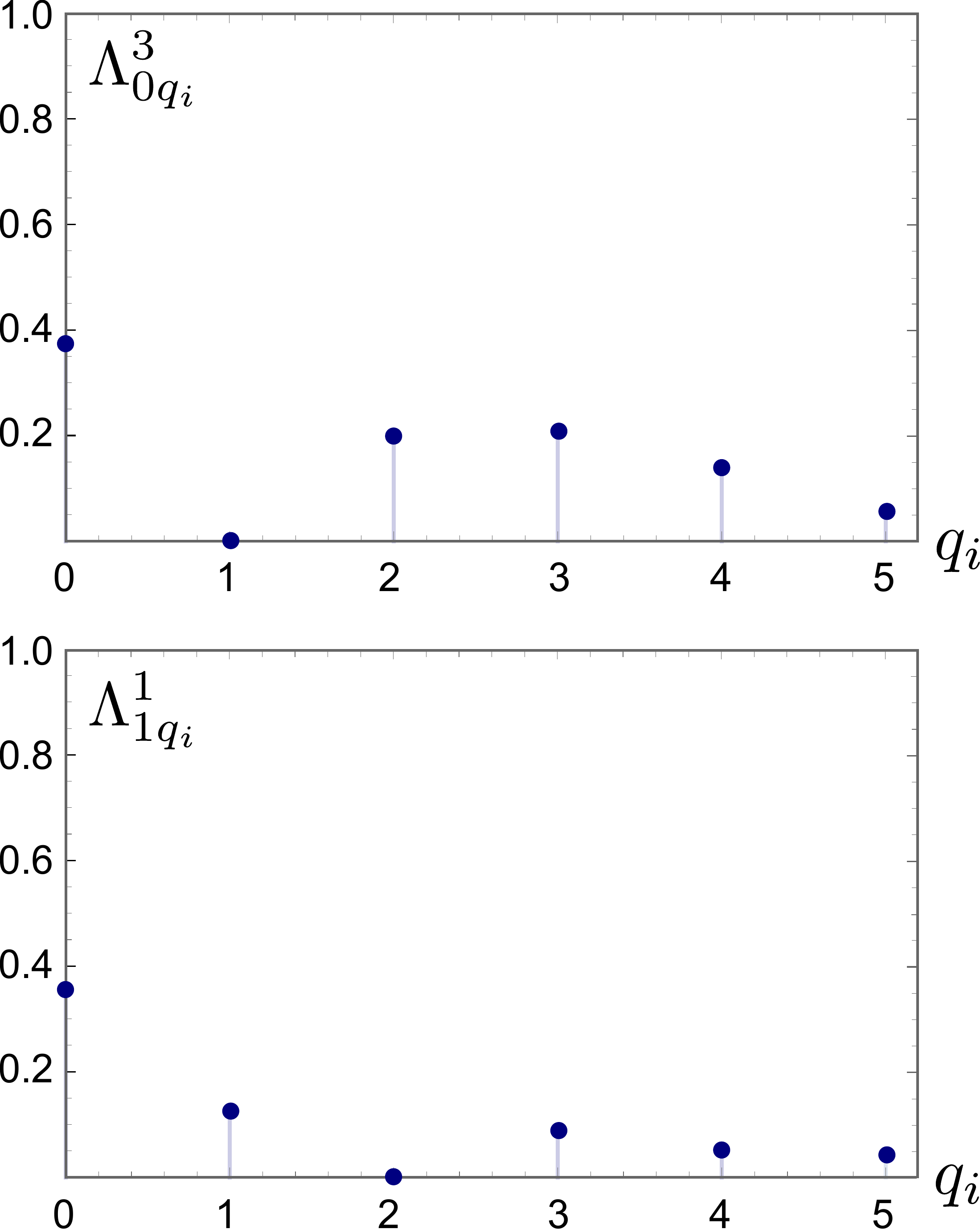}
	\caption{Pump-signal-idler spatial overlap for $l=\pm 3$ (top) and $l=\pm 1$ (bottom) as a function of the idler radial orders, when the pump parameters are fixed at $l_p = 0$ and $q_p = 2\,$.}
	\label{fig:overlap_3rd_order}
\end{figure}

The simplest case with more than one Poincar\'e sphere symmetry is realized by a third-order seed beam. 
As before, the pump is assumed to carry zero OAM and optimal overlap is attained with $q_p = 2\,$. 
In Fig.\ref{fig:overlap_3rd_order} we show the overlap integrals for different OAM values. As we 
can see, the largest coupling between signal and idler modes with $l=\pm 1$ and $\pm3$ occurs for 
the idler radial index $q_i = 0\,$. Therefore, the $l=\pm 3$ sphere represents signal and idler modes 
with $q_s = q_i = 0\,$. However, the $l=\pm 1$ sphere represents signal modes with $q_s = 1$ and 
idler modes with $q_i = 0\,$. In any case, the radial numbers are irrelevant for the Poincar\'e 
symmetry condition, which is essentially determined by OAM conservation.

\begin{figure}[h!]
	\includegraphics[scale=0.17]{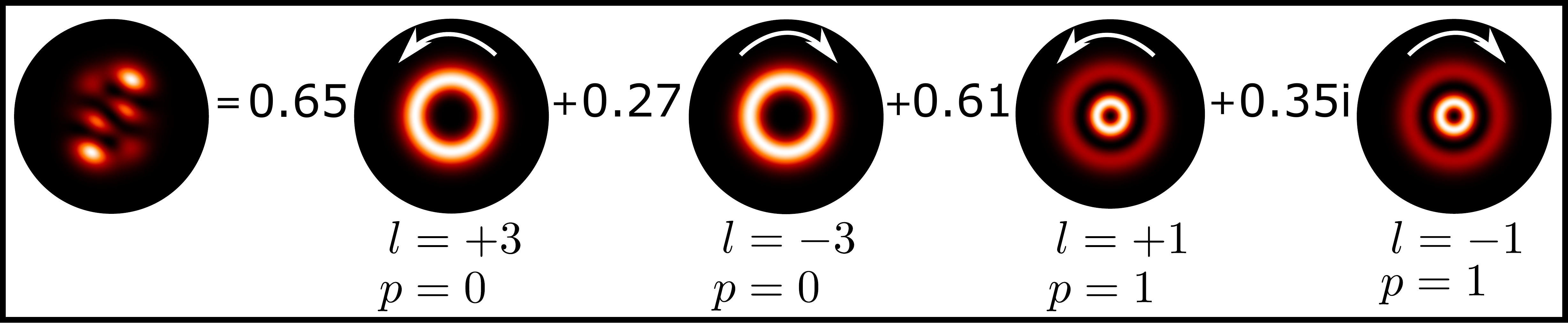}
	\caption{Example of a third-order seed beam with $A_{in}^{1 1}=A_{in}^{3 0}=1/\sqrt{2}\,$. The coordinates on the Poincar\'e spheres for $l=\pm 3$ and 	$l=\pm 1$ are ($\theta_3=45^o$, $\phi_3=0^o$) and ($\theta_1=60^o$, $\phi_1=90^o$), respectively.}
	\label{fig:third_seed}
\end{figure}

With these input modes, the source terms in the dynamical equations become
\begin{eqnarray}
    \!\!\!\!\!\!\!\!\!
    \mathcal{E}_{p_{in}} &=& \alpha_{p_{in}} \,\psi^{0 2}\,,
    \nonumber\\
    \!\!\!\!\!\!\!\!\!
    \mathcal{E}_{s_{in}} &=& A_{in}^{3 0} 
    \left[\cos\left(\theta_3/2\right) \psi^{3 0} + e^{i\phi_3} \sin\left(\theta_3/2\right) \psi^{-3 0}\right]
    \nonumber\\
    \!\!\!\!\!\!\!\!\!
    &+& A_{in}^{1 1} 
    \left[\cos\left(\theta_1/2\right) \psi^{1 1} + e^{i\phi_1} \sin\left(\theta_1/2\right) \psi^{-1 1}\right]\,.
    \label{input-3rd}
\end{eqnarray}
Two sets of Poincar\'e sphere coordinates are used, $(\theta_1,\phi_1)$ and 
$(\theta_3,\phi_3)\,$, associated with the $l=\pm 1$ and $l=\pm 3$ manifolds, respectively.
In Fig. \ref{fig:third_seed} we show an example of seed beam and its decomposition in these
manifolds.

The steady state intracavity pump amplitude is given by the solution of the quintic equation
\begin{equation}
\left[ \gamma_r +
\frac{\abs{\Lambda^3_{00}\,A_{in}^{30}}^2}{\left(1-\abs{\Lambda^3_{00}\,\beta_p}^2\right)^2}
+
\frac{\abs{\Lambda^1_{10}\,A_{in}^{11}}^2}{\left(1-\abs{\Lambda^1_{10}\,\beta_p}^2\right)^2}
\right] \beta_p = \beta_{p_{in}}\,.  
\label{eq:ss-pump-3rd-order}
\end{equation}
Once the solution of Eq. \eqref{eq:ss-pump-3rd-order} is obtained, the intracavity signal and idler spatial structures can be readily calculated from
\begin{eqnarray}
    \!\!\!\!\!\!\!\!\!
    \mathcal{E}_{s} &=& 
    \xi_{s}^{3 0} 
    \left[\cos\left(\theta_3/2\right) \psi^{3 0} + e^{i\phi_3} \sin\left(\theta_3/2\right) \psi^{-3 0}\right],
    \nonumber\\
    \!\!\!\!\!\!\!\!\!
    &+&
    \xi_{s}^{1 1} 
    \left[\cos\left(\theta_1/2\right) \psi^{1 1} + e^{i\phi_1} \sin\left(\theta_1/2\right) \psi^{-1 1}\right],
    \nonumber\\
    \!\!\!\!\!\!\!\!\!
    \mathcal{E}_{i} &=& 
    \xi_{i}^{3 0} 
    \left[e^{-i\phi_3} \sin\left(\theta_3/2\right) \psi^{3 0} + \cos\left(\theta_3/2\right) \psi^{-3 0}\right],
    \nonumber\\
    \!\!\!\!\!\!\!\!\!
    &+&
    \xi_{i}^{1 0} 
    \left[e^{-i\phi_1} \sin\left(\theta_1/2\right) \psi^{1 0} + \cos\left(\theta_1/2\right) \psi^{-1 0}\right],
    \label{signal-idler-intracavity-3rd-order}
\end{eqnarray}
with the mode amplitudes given by
\begin{eqnarray}
    \xi_{s}^{3 0} &=& \frac{\eta\,A_{in}^{3 0}/\gamma}{1-\abs{\Lambda^{3}_{0 0} \beta_p}^2} , 
    \quad 
    \xi_{s}^{1 1} = \frac{\eta\,A_{in}^{1 1}/\gamma}{1-\abs{\Lambda^{1}_{1 0} \beta_p}^2} , 
    \nonumber\\
    \xi_{i}^{3 0} &=& i \beta_p\,\Lambda^{3\,\ast}_{0 0}\, \left(\xi_{s}^{3 0}\right)^\ast , 
    \; 
    \xi_{i}^{1 0} = i \beta_p\,\Lambda^{1\,\ast}_{1 0}\, \left(\xi_{s}^{1 1}\right)^\ast\,.
    \label{signal-idler-csi-3rd-order}
\end{eqnarray}
\begin{figure}[h!]
\includegraphics[scale=0.087]{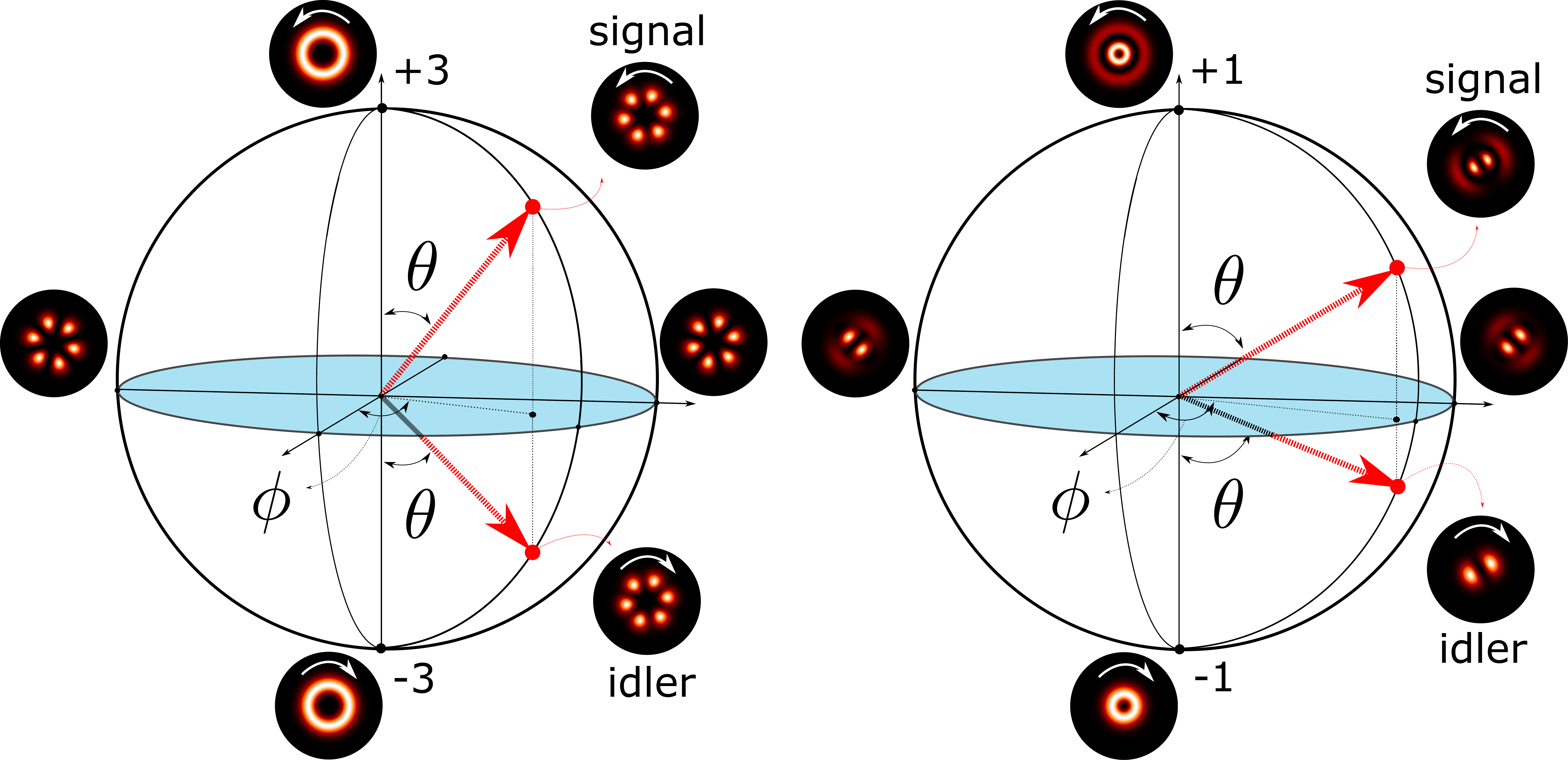}
\caption{Third-order OAM symmetry. On left side we can visualize the symmetry relation verified in the Poincar\'e 
sphere for $l=\pm 3$ and on the right side the same symmetry is displayed on the independent sphere for $l=\pm 1\,$.}
\label{fig:esfera_3}
\end{figure}
\begin{figure}[h!]
\includegraphics[scale=0.205]{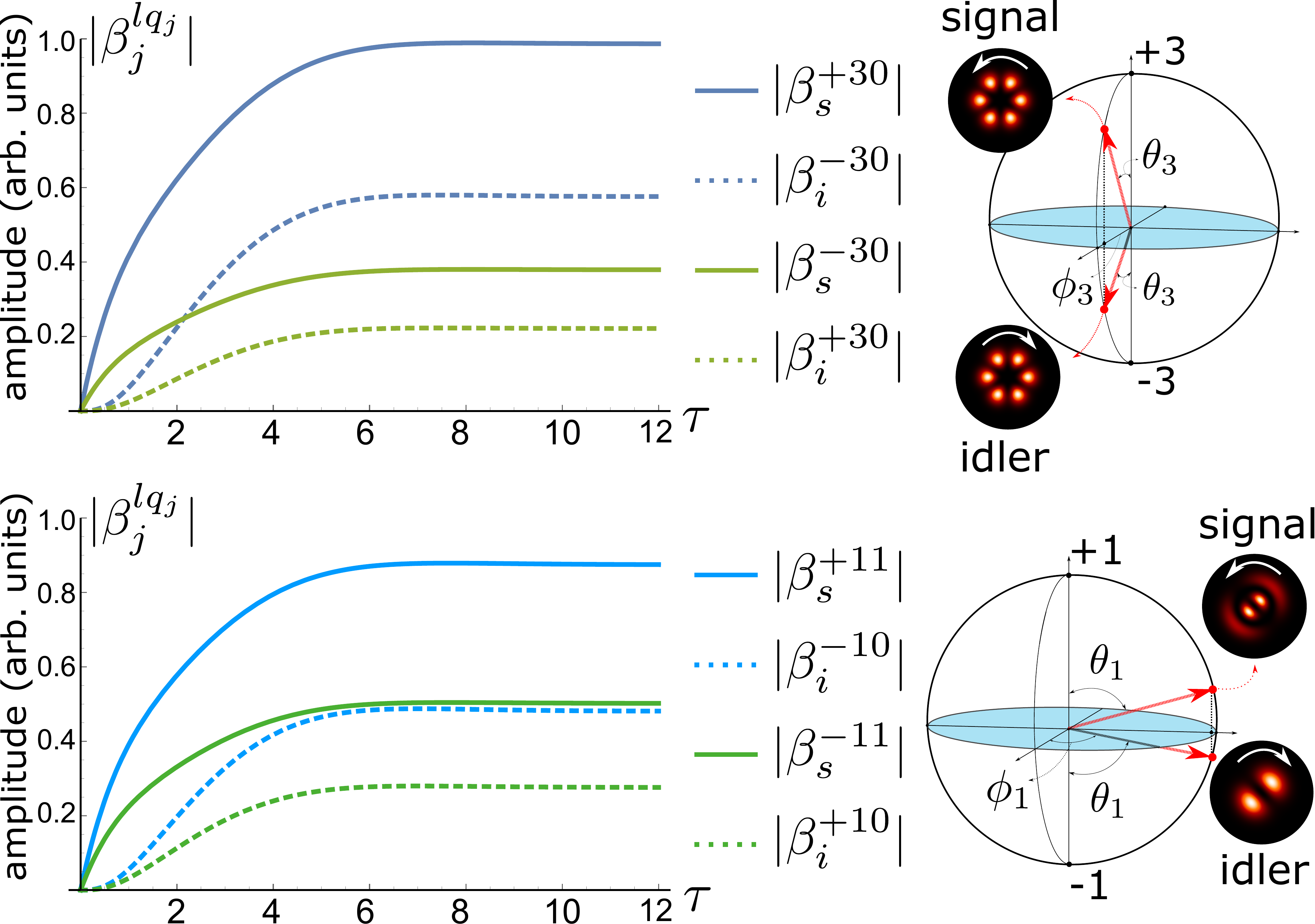}
\caption{Two-sphere symmetry for the numerical simulation of the OPO dynamics under third-order injection. 
The spherical coordinates for $l=\pm 3$ are $\theta_3 = 0.73\,\mathrm{rad}$ and 
$\phi_3 = 0\,$. For $l=\pm 1$ the coordinates are $\theta_1 = 1.04\,\mathrm{rad}$ and 
$\phi_1 = 1.57\,\mathrm{rad}\,$. The pump, seed and decay parameters were set to 
$\beta_{p_{in}} = 2\,$, $A^{3 0}_{in} = A^{1 1}_{in} = 1/\sqrt{2}$ and $\gamma_r = 1\,$, 
respectively. The free-running oscillation threshold is $\beta_{p_{th}} \approx 2.7\,$.}
\label{fig:TimeEvolutionThirdOrder1}
\end{figure}
These mode superpositions are represented in the Poincar\'e spheres shown if Fig. \ref{fig:esfera_3}. The signal and idler coordinates are related by
\begin{eqnarray}
    \theta_3^i &=& \pi - \theta_3^s\,,\qquad \phi_3^i = \phi_3^s\;,
    \nonumber\\
    \theta_1^i &=& \pi - \theta_1^s\,,\qquad \phi_1^i = \phi_1^s\;.
\end{eqnarray}
These relations show that the signal and idler modes verify the Poincar\'e symmetry independently on each 
sphere. We have also tested the symmetry condition with a numerical integration of the dynamical equations. 
The numerical results for the signal and idler mode amplitudes are displayed in 
Fig. \ref{fig:TimeEvolutionThirdOrder1}, along with the resulting two-sphere representation. The spherical 
coordinates are evaluated from the numerical results, confirming the two-sphere symmetry condition.

As before, this condition ensures both maximal intensity overlap and OAM conservation.
The extension of this symmetry condition to higher orders is straightforward. The mode space can be split in 
two-dimensional OAM subspaces with fixed absolute value $\abs{l} = 1,2,3...$, where the symmetry condition 
is simultaneously verified on independent Poincar\'e spheres. For even orders, there is an isolated component 
with zero OAM, where no symmetry condition can be realized other than the usual phase conjugation between signal 
and idler amplitudes.

\section{OAM-structured Pump}
\label{sec:structured-pump}

\begin{figure}[h!]
\includegraphics[scale=0.09]{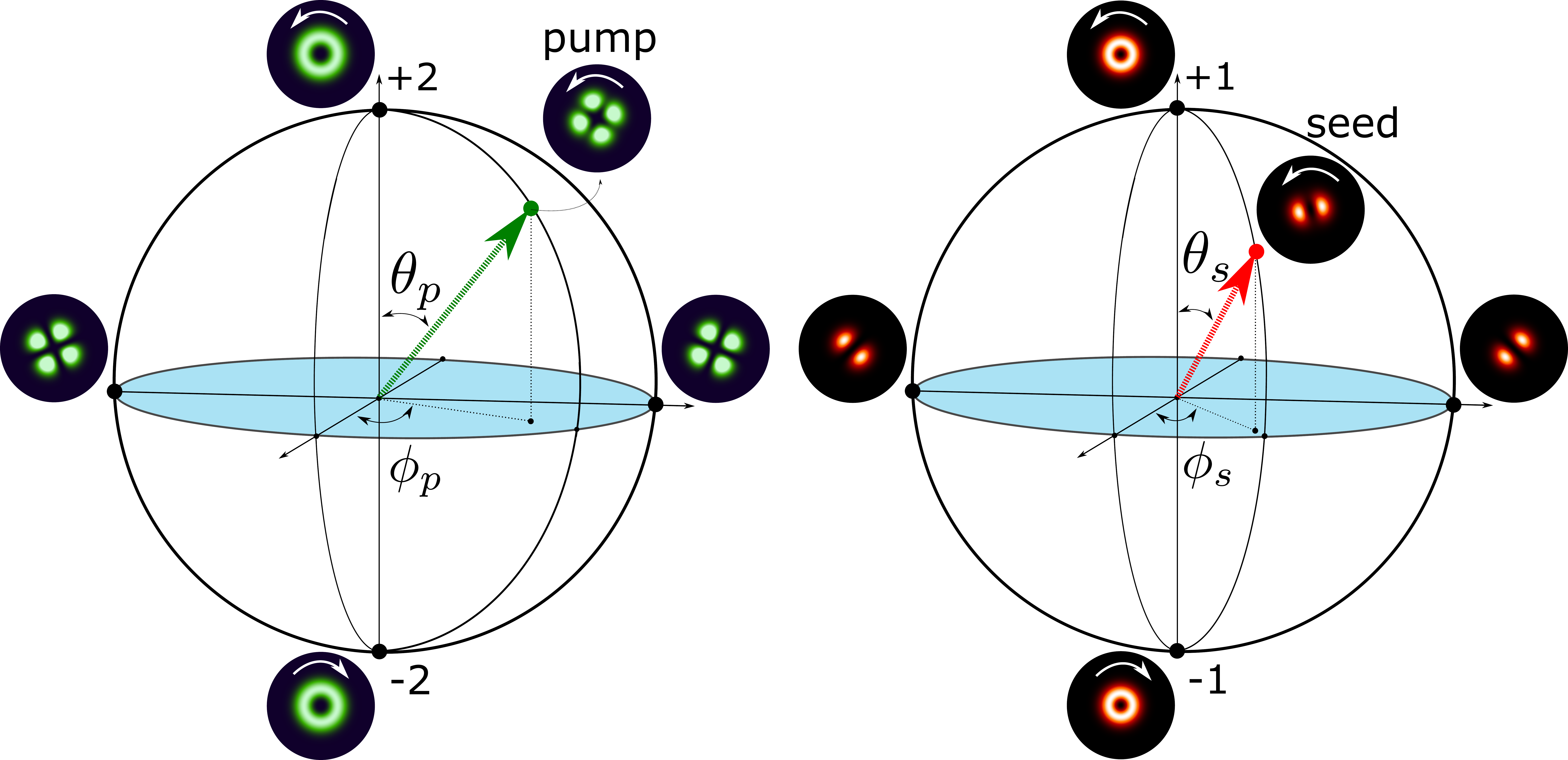}
\caption{Poincar\' e sphere representations of the pump ($l=\pm 2$) and seed ($l=\pm 1$) beams.}
\label{fig:esfera_pump_seed}
\end{figure}

It is also interesting to investigate how the Poincar\'e sphere symmetry between signal and idler is affected by a 
pump beam carrying OAM. In this case, a Poincar\'e representation also applies to the pump beam. For example, 
let us consider a pump beam prepared in a superposition of second-order modes with $l=\pm 2$ and $q_p=0\,$. 
The seed beam is assumed to be a first-order superposition of modes with $l=\pm 1\,$. 
The pump and seed input amplitudes can be written as 
\begin{eqnarray}
\!\!\!\!\!\!\!\!\!
\mathcal{E}_{p_{in}} \!&=&\! A_{p_{in}}
\left[\cos\left(\theta_p/2\right) \psi^{2 0} + e^{i\phi_p} \sin\left(\theta_p/2\right) \psi^{-2 0}\right],
\nonumber\\
\!\!\!\!\!\!\!\!\!
\mathcal{E}_{s_{in}} \!&=&\! A_{s_{in}}
\left[\cos\left(\theta_s/2\right) \psi^{1 0} + e^{i\phi_s} \sin\left(\theta_s/2\right) \psi^{-1 0}\right],
\label{eq:OAM-pump-input}
\end{eqnarray}
where $(\theta_p,\phi_p)$ and $(\theta_s,\phi_s)$ are the Poincar\'e sphere coordinates of the input pump and seed beams,
respectively. In Fig. \ref{fig:esfera_pump_seed}, we display the Poincar\'e representation of the pump (left) and seed (right) input modes. 
The idler modes with optimal intensity overlap and OAM conservation with the seed are also first order LG modes with 
$l=\pm 1$ and $q_i = 0\,$. 

The time evolution of the pump, signal and idler intracavity amplitudes is governed by the following dynamical equations
\begin{eqnarray}
\!\!\!\!\!\!\!\!\!
\dot{\beta}^{+}_{p} &=& - \gamma_r \beta^{+}_p + i\Lambda\, \beta^{+}_s\beta^{+}_i + 
\beta_{p_{in}} \cos\left(\theta_p/2\right) ,
\nonumber\\
\!\!\!\!\!\!\!\!\!
\dot{\beta}^{-}_{p} &=& - \gamma_r \beta^{-}_p + i\Lambda\, \beta^{-}_s\beta^{-}_i + 
\beta_{p_{in}} e^{i\phi_p}\sin\left(\theta_p/2\right) ,
\nonumber\\
\!\!\!\!\!\!\!\!\!
\dot{\beta}^{+}_{s} &=& - \beta^{+}_s + i\Lambda^\ast\, \beta^{+}_p\beta^{+\,*}_i + 
\beta_{s_{in}} \cos\left(\theta_s/2\right) ,
\label{eq:dyneq-OAM-pump}\\
\!\!\!\!\!\!\!\!\!
\dot{\beta}^{-}_{s} &=& - \beta^{-}_s + i\Lambda^\ast\, \beta^{-}_p\beta^{-\,*}_i + 
\beta_{s_{in}} e^{i\phi_s}\sin\left(\theta_s/2\right) ,
\nonumber\\
\!\!\!\!\!\!\!\!\!
\dot{\beta}^{\pm}_{i} &=& - \beta^{\pm}_i + i\Lambda^\ast\, \beta^{\pm}_p\left(\beta^{\pm}_s\right)^\ast , 
\nonumber
\end{eqnarray}
where $\beta_p^{\pm}$ are the normalized pump amplitudes for $l=\pm 2\,$, $\beta_{s(i)}^{\pm}$,are the normalized 
seed (idler) amplitudes for $l=\pm 1\,$, and the time derivatives in the left-hand-side are taken with respect to 
the dimensionless parameter $\tau = \gamma t\,$. The mode coupling is mediated by the three-mode overlap
\begin{eqnarray}
\Lambda = \int \left[\psi_p^{\pm 2 0}(\mathbf{r})\right]^\ast\,
\psi_s^{\pm 1 0}(\mathbf{r})\,\psi_i^{\pm 1 0}(\mathbf{r})\,d^2\mathbf{r}\;.
\label{eq:overlap-OAM-pump}
\end{eqnarray}
The steady state solutions are obtained by setting the time derivatives equal to zero 
in the left-hand-side of Eqs.\eqref{eq:dyneq-OAM-pump} and solving the resulting algebraic equations. 
From the last three equations in \eqref{eq:dyneq-OAM-pump}, we have
\begin{eqnarray}
\!\!\!\!\!\!\!\!
\beta^{+}_s &=& \xi^+_s\,\cos\left(\theta_s/2\right), 
\quad
\beta^{-}_s =  \xi^-_s\,e^{i\phi_s}\sin\left(\theta_s/2\right),
\nonumber\\
\!\!\!\!
\beta^{+}_i &=&  \xi^+_i\,\cos\left(\theta_s/2\right),
\quad
\beta^{-}_i =  \xi^-_i\,e^{-i\phi_s}\sin\left(\theta_s/2\right),
\label{eq:ss-signal-idler-OAM-pump}
\end{eqnarray}
where we defined 
\begin{eqnarray}
\xi^{\pm}_s &=& \frac{\eta\,A_{s_{in}}/\gamma}{1-\abs{\Lambda\,\beta^{\pm}_p}^2}\,,
\nonumber\\
\xi^{\pm}_i &=& i\beta_p^{\pm}\Lambda^{\ast}\left(\xi^{\pm}_s\right)^{\ast}
\end{eqnarray}

The steady state intracavity pump amplitudes are given by the solutions of two independent quintic equations
\begin{eqnarray}
&&\left[\gamma_r + \frac{\abs{\Lambda\,A_{s_{in}}}^2 \cos^2\left(\frac{\theta_s}{2}\right)}
{\left(1-\abs{\Lambda\,\beta^+_p}^2\right)^2}\right]\!\!\beta^+_p 
= \beta_{p_{in}} \!\cos\left(\frac{\theta_p}{2}\right) ,
\label{eq:ss-pump-OAM-pump}\\
&&\left[\gamma_r + \frac{\abs{\Lambda\,A_{s_{in}}}^2 \sin^2\left(\frac{\theta_s}{2}\right)}
{\left(1-\abs{\Lambda\,\beta^-_p}^2\right)^2}\right]\!\!\beta^-_p =
\beta_{p_{in}} e^{i\phi_p} \sin\left(\frac{\theta_p}{2}\right) .
\nonumber
\end{eqnarray}
From Eqs. \eqref{eq:ss-signal-idler-OAM-pump} we immediately obtain the complete structure of the 
intracavity signal and idler fields as
\begin{eqnarray}
\!\!\!\!\!\!\!\!\!\!
    \mathcal{E}_{s} &=& \xi^+_{s} \cos\left(\theta_{s}/2\right) \psi_s^{1 0} + 
    \xi^-_{s} e^{i\phi_s} \sin\left(\theta_s/2\right) \psi_s^{-1 0}\,,
    \nonumber\\
\!\!\!\!\!\!\!\!\!\!
    \mathcal{E}_{i} &=& \xi^+_{i} \cos\left(\theta_{s}/2\right) \psi_i^{1 0} + 
    \xi^-_{i} e^{-i\phi_s} \sin\left(\theta_s/2\right) \psi_i^{-1 0} .
    \label{signal-idler-intracavity-OAM-pump}
\end{eqnarray}
The Poincar\'e sphere representation of signal and idler depends on both the seed and 
pump parameters. While the seed parameters are explicit in the expressions above, the dependence on 
the pump is implicit through $\xi^{\pm}_{s,i}\,$. 

Assuming weak pump and seed powers, the pump is not significantly depleted by the parametric 
interaction and the intracavity pump is essentially driven by the external input, leading to
\begin{eqnarray}
\beta^{+}_p &\approx& \frac{\beta_{p_{in}}}{\gamma_r} \cos\left(\theta_{p}/2\right) \,, 
\nonumber\\
\beta^{-}_p &\approx& \frac{\beta_{p_{in}}}{\gamma_r} e^{i\phi_p}\sin\left(\theta_{p}/2\right) \,. 
\label{undepleted_cond}
\end{eqnarray}
In this case, we can write the analytical solutions for signal and idler as
\begin{eqnarray}
\!\!\!\!\!\!
\mathcal{E}_s &=& \xi_s^+ \cos{\left(\frac{\theta_s}{2}\right)}\psi^{+10} +\xi_s^- e^{i\phi_s}\sin{\left(\frac{\theta_s}{2}\right)}\psi^{-10} ,
\label{signal_ordem_1}  \\
\!\!\!\!\!\!
\mathcal{E}_i &=& \xi_i^+ \cos{\left(\frac{\theta_s}{2}\right)}\cos{\left(\frac{\theta_p}{2}\right)}\psi^{+10} 
\nonumber\\
\!\!\!\!\!\!
&+& \xi_i^- e^{i(\phi_p - \phi_s)}\sin{\left(\frac{\theta_s}{2}\right)}\sin{\left(\frac{\theta_p}{2}\right)}\psi^{-10} \,, \label{idler_ordem_1}
\end{eqnarray}
where the amplitudes can be explicitly written in terms of the pump parameters as
\begin{eqnarray}
\xi_s^{\pm} &=& \frac{\eta A_{s_{in}}/\gamma}{1-\frac{\abs{\Lambda\beta_{p_{in}}}^2}{2\gamma_r^2}(1\pm\cos\theta_p)}\,,
\nonumber\\
\xi_i^\pm &=& i\beta_p^{\pm}\Lambda^{\ast}\left(\xi^{\pm}_s\right)^{\ast}
\end{eqnarray}

\begin{figure}[h]
\includegraphics[scale=0.12]{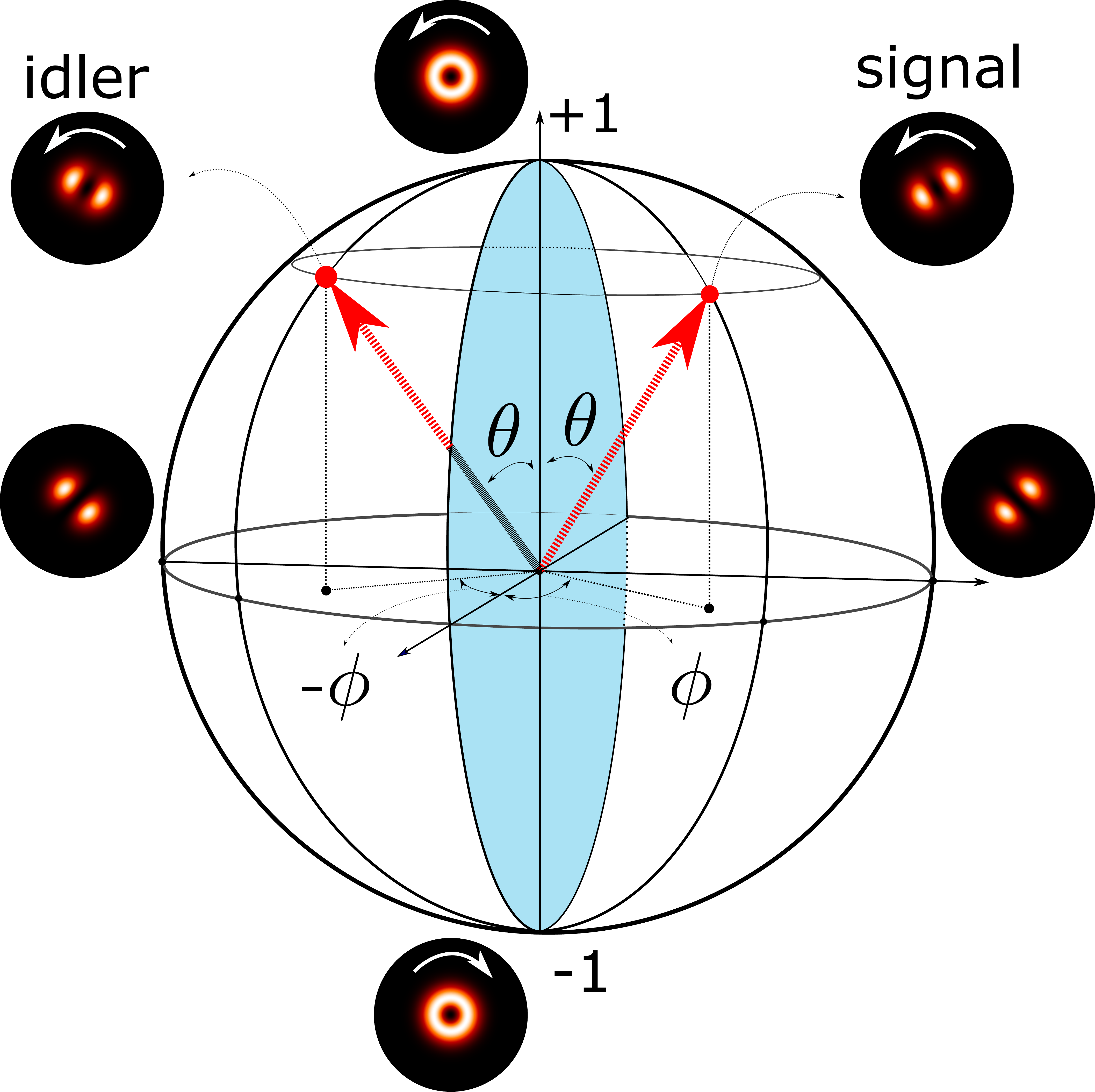}
\caption{Poincaré sphere representation of the intracavity signal and idler structures when $\theta_p = \pi/2$ and 
$\phi_p = 0\,$.}
\label{fig:simetria_pump_ordem2}
\end{figure}

In this regime, the Poincar\'e symmetry becomes more evident for special values of the pump and seed 
parameters. First, for $\theta_p = \pi/2$ and arbitrary $\phi_p\,$, we easily obtain that $\theta_i = \theta_s$ 
and $\phi_i = \phi_p - \phi_s\,$. This symmetry is displayed in Fig. \ref{fig:simetria_pump_ordem2} for 
$\phi_p = 0\,$. The idler spatial structure is represented by the specular image of the signal with respect 
to the great circle $\phi = 0\,$.

Moreover, restricting the input pump power to values well below the free-running oscillation threshold,
such that
\begin{equation}
\frac{\abs{\Lambda\beta_{p_{in}}}^2}{2\gamma_r^2} \ll 1\,,
\label{threshold_cond_OPO}
\end{equation}
one can take $\xi_s^{+}\approx\xi_s^{-}$ and $\xi_i^{+}\approx\xi_i^{-}\,$. From 
Eqs. \eqref{signal_ordem_1} and \eqref{idler_ordem_1}, it is easy to see that the idler parameters are 
related to the pump and seed parameters as
\begin{eqnarray}
\phi_i&=&\phi_p-\phi_s\,, \\
\tan\left(\frac{\theta_i}{2}\right)&=&
\tan\left(\frac{\theta_p}{2}\right)\tan\left(\frac{\theta_s}{2}\right)\,.
\end{eqnarray}
Therefore, the seed and idler beams are azimutally symmetric in the first order Poincar\'e sphere with respect to the the angle $\phi_p/2$, while the idler polar location follows a nontrivial relation with the pump and seed polar parameters. Another interesting condition within this regime occurs for $\theta_s = \pi/2$ and $\phi_s = 0\,$. 
In this case, the idler parameters become 
equal to those of the pump, $\theta_i = \theta_p$ and $\phi_i = \phi_p\,$. Therefore, the idler spatial structure can be 
actively controlled by fixing either the pump or the seed parameters and varying the others. In a real experimental 
situation, this active control can be challenged by the astigmatic effects caused by the crystal birrefringence 
\cite{martinelli2004}. However, this effect can be compensated for with a two-crystal setup, as the one used in 
Ref. \cite{gao2014,gao2018}. Active control of signal and idler spatial structures exploring symmetry conditions 
can be useful for shaping spatial quantum correlations generated in parametric amplification.

\section*{Acknowledgments}

Funding was provided by 
Conselho Nacional de Desenvolvimento Cient\'{\i}fico e Tecnol\'ogico (CNPq),
Coordena\c c\~{a}o de Aperfei\c coamento de Pessoal de N\'\i vel Superior (CAPES), 
Funda\c c\~{a}o Carlos Chagas Filho de Amparo \`{a} Pesquisa do Estado do Rio de Janeiro (FAPERJ), 
and Instituto Nacional de Ci\^encia e Tecnologia de Informa\c c\~ao Qu\^antica 
(INCT-IQ 465469/2014-0).

\bibliography{bibliography.bib}

\end{document}